\documentclass[hypertexnames=false,superscriptaddress,twocolumn,aps,prl]{revtex4-2}
\usepackage{array, makecell}

\usepackage{dcolumn}
\usepackage{bm}
\usepackage{tikz}
\usepackage{pgfplots}
\pgfplotsset{compat=1.18}
\tikzset{>=latex}
\usepackage{amsfonts}

\usepackage{booktabs}

\usepackage{lipsum}
\usepackage{graphicx} 
\usepackage{standalone}
\usepackage[caption=false]{subfig}
\usepackage[percent]{overpic}
\usepackage{verbatim}
\usepackage{tcolorbox}

\usepackage{physics} 
\usepackage{enumerate}
\usepackage{dsfont}
\usepackage{xcolor}
\usepackage{lipsum}
\usepackage[normalem]{ulem}
\usepackage{hyperref}
\usepackage[capitalise]{cleveref}

\begin{document}

\title{Addressable gate-based logical computation with quantum LDPC codes}

\author{Laura Pecorari}
\thanks{These authors contributed equally to this work.}
\affiliation{University of Strasbourg and CNRS, CESQ and ISIS (UMR 7006), aQCess, 67000 Strasbourg, France}
\author{Francesco Paolo Guerci}
\thanks{These authors contributed equally to this work.}
\affiliation{University of Strasbourg and CNRS, CESQ and ISIS (UMR 7006), aQCess, 67000 Strasbourg, France}
\author{Hugo Perrin}
\thanks{These authors contributed equally to this work.}
\affiliation{University of Strasbourg and CNRS, CESQ and ISIS (UMR 7006), aQCess, 67000 Strasbourg, France}
\affiliation{QPerfect, 23 Rue du Loess, 67200 Strasbourg, France}
\author{Guido Pupillo}
\email{pupillo@unistra.fr}
\affiliation{University of Strasbourg and CNRS, CESQ and ISIS (UMR 7006), aQCess, 67000 Strasbourg, France}
\affiliation{Institut Universitaire de France (IUF), 75000 Paris, France}

\date{\today}

\begin{abstract}

Quantum computing relies on quantum error correction for high-fidelity logical operations, but scaling to achieve near-term quantum utility is highly resource-intensive. High-rate quantum LDPC codes can reduce error correction overhead, yet realizing high-rate fault-tolerant computation with these codes remains a central challenge. Lattice surgery remains the main approach for addressable logical computation but at a cost of $\mathcal{O}(d)$ time overhead. Other approaches rely on low-rate code switching methods or high space overhead homomorphic CNOT. In this work, we introduce a gate-based protocol for addressable single- and multi-qubit Clifford operations on individual logical qubits encoded within one or more quantum LDPC codes. Our scheme leverages logical transversal operations via an auxiliary -- a Bacon-Shor code or an unrotated surface code -- to perform logical operations with constant time overhead enabled by teleportation. We demonstrate the implementation of an overcomplete logical Clifford gate set and perform numerical simulations to evaluate the error-correction performance of our protocol. Additionally, our scheme can also realize any joint Pauli product measurement natively. Finally, we observe that our scheme can be integrated with magic state cultivation protocols to achieve universal, gate-based, and fully addressable quantum computation. 
\end{abstract}

\maketitle
\def\thefootnote{*}\footnotetext{These authors contributed equally to this work.}

Executing parallel logical operations on  error-corrected logical qubits is key to enable quantum computation. Quantum error correction (QEC) uses redundancy to delocalize the state of individual logical qubits over several physical qubits to increase protection against errors \cite{gottesman1997stabilizercodesquantumerror}. This comes at the price of large qubit and gate overheads, which ultimately make QEC highly resource-intensive and pose challenges to scalability. 

High-rate quantum Low-Density Parity-Check (LDPC) codes \cite{Breuckmann_2021} offer a promising route toward low-overhead, error-corrected quantum memories. By exploiting long-range connectivity, these codes can compactly encode a large number of logical qubits within a single quantum register and have been shown to outperform standard locally connected codes, such as surface codes \cite{Kitaev_2003,Dennis_2002}, under circuit-level noise models \cite{Bravyi2024,xu2023constantoverhead,pecorari2024highratequantumldpccodes}. However, this dense information encoding poses a major challenge for realizing high-rate quantum computation on near-term quantum processors.

The standard theoretical framework to date for performing quantum computation with high-rate quantum LDPC codes uses lattice surgery to measure logical Pauli operators \cite{Horsman2012,Cohen2022,Cross2024,cowtan2024ssipautomatedsurgeryquantum,Williamson2024,Cowtan2025,Ide2025,Swaroop2025,Zhang_2025,yoder2025tourgrossmodularquantum}. This paradigm consists in filling the region between distant code patches with additional ancilla qubits to enable local connectivity and entangle distant logical qubits by measuring the ancilla qubits. This comes at the price of a $\mathcal{O}(d)$ time overhead and a $\mathcal{O}(d^2)$ or $\mathcal{O}(d)$ space -- i.e. qubit -- overhead, depending on the protocols, for a distance $d$ quantum LDPC code. The lattice surgery approach is particularly favorable for superconducting platforms, because of the fast measurements and native local connectivity. It is however inconvenient for other platforms such as neutral atoms and trapped ions, where measurements are typically the slowest operations \cite{PhysRevLett.105.100501,doi:10.1126/sciadv.adv2590,bluvstein2025architecturalmechanismsuniversalfaulttolerant}. 
Alternative approaches for gate-based computing are instead based on code-switching from an LDPC code to lower-rate codes such as the surface code \cite{xu2023constantoverhead} or on performing global transversal operations on all logical qubits at the same time \cite{JiangPhysRevX.15.021065}. However, code-switching results in large qubit overheads -- thus reducing the appeal of LDPC memories --, while restriction to global logical operations is incompatible with efficient compilation of quantum algorithms. For the latter, it would be highly desirable to invent protocols to perform logical operations on individual logical qubits within a given LDPC code or across different LDPC codes.

\begin{figure*}[t!]
    \centering
    \includegraphics[width=0.99\linewidth]{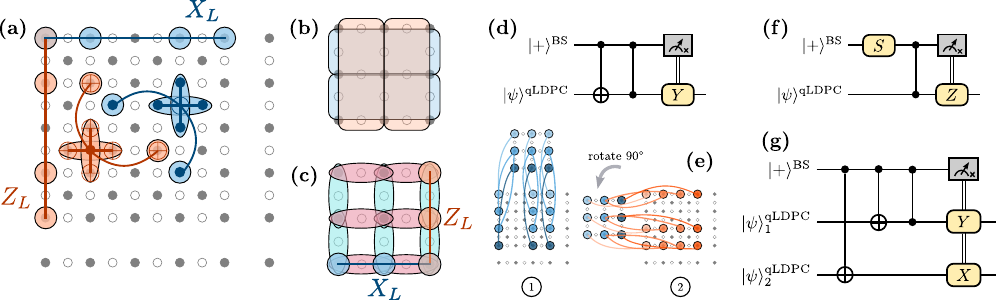}
    \caption{(a) Example of La-cross code with next-to-nearest neighbor connectivity (labeled as $k=2$). The code is defined by a $n\times n$ lattice with a $(n-k)\times(n-k)$ sublattice with open boundary conditions. A $X$- (blue) and a $Z$-stabilizer (red) are depicted, along with one pair of logical $X_L$ (blue) and $Z_L$ (red) operators. (b) Bacon-Shor code with non-local $X$- (blue) and $Z$-stabilizers (red) spanning along two consecutive rows and columns of the array, respectively. (c) $X$-type (cyan) and $Z$-type (pink) two-body gauge operators of the Bacon-Shor code and its two logical $X_L$ (blue) and $Z_L$ (red) operators. (d) Circuit implementing a logical $\mathcal{R}_y(\pi/2)$ rotation on a single quantum LDPC logical qubit via teleportation and (e) example of implementation via qubit shuttling. Different equivalent representatives of the same logical operator are colored with different gradients of the same color (blue for $X_L$ and red for $Z_L$). (f) Circuit implementing a logical S gate via teleportation. (g) Circuit implementing a logical two-qubit entangling operation between two logical qubits via teleportation.}
    \label{fig:fig1}
\end{figure*}

In this work, we propose a gate-based protocol to realize an overcomplete set of logical single- and multi-qubit Clifford gates on individual logical qubits within a LDPC code or across different LDPC codes. Our scheme employs transversal entangling gates between a subset of physical qubits of the quantum LDPC code and all physical qubits of an auxiliary smaller code, a Bacon–Shor code \cite{baconPhysRevA.73.012340} or an unrotated surface code in this work, enabling parallel and fully addressable logical operations via teleportation.
As an example, we present supporting numerical simulations for both Bacon-Shor and surface code \emph{gadgets} to show the error correction performance of our of the full logical and addressable Clifford logical gate set compared to a memory experiment on the same code. 
As evidence of the versatility of our protocol, we also discuss how the logical Clifford gate set enabled by our gadget can be augmented with the Pauli-based computation framework by performing joint measurements of arbitrary logical operators of the targeted code.
Finally, we briefly comment on how to integrate our protocol with magic state cultivation methods \cite{gidney2024magicstatecultivationgrowing,vaknin2025efficientmagicstatecultivation,claes2025cultivatingtstatessurface,sahay2025foldtransversalsurfacecodecultivation} to enable universal addressable quantum computation in the quantum LDPC code. Although the proposed protocol targets the La-cross quantum LDPC code family, we detail the necessary and sufficient conditions for our logical gadget and outline an approach to generalize this gate-based framework to other quantum LDPC codes, including constant-rate families.

Our protocol requires a $\mathcal{O}(d^2)$ qubit overhead with only a $\mathcal{O}(1)$ time overhead \cite{Zhou_2025} for a distance $d$ quantum LDPC code, independent of whether the logical gate is single- or multi-qubit. Hence, it outperforms known surgery-based schemes in terms of time overhead, while offering a comparable or better spacetime overhead. Our protocol is well-suited for near-term implementation on platforms with non-local connectivity, such as neutral atoms \cite{bluvstein2025architecturalmechanismsuniversalfaulttolerant} or trapped ions \cite{ions2002}, using qubit shuttling \cite{Bluvstein_2022}, atomic buses \cite{bergonzoni2026faulttolerantquantumcomputationstatic}, native long-range Rydberg \cite{pecorari2024highratequantumldpccodes,poole2024architecturefastimplementationqldpc,saffman2025quantumcomputingatomicqubit} or Coulomb interactions, respectively. However, the protocol is hardware-agnostic and can be applied to other platforms with sufficient long-range connectivity. In particular, superconducting qubits with non-local couplers \cite{PhysRevLett.129.050504,PRXQuantum.6.010306,yoder2025tourgrossmodularquantum} are also a promising candidate to perform, e.g., Pauli-based computation with our gadget.

\section{Results}
\subsection{La-cross codes and structure of logical operators}
We consider the Calderbank-Shor-Steane (CSS) \cite{Calderbank_1996,Steane1996} La-cross code family of high-rate quantum LDPC codes, which has recently been shown to outperform the surface code in experimentally relevant regimes of physical error probability with different noise models \cite{pecorari2024highratequantumldpccodes,pecorari2025quantumldpccodeserasurebiased}. This choice is motivated by two key reasons. First, La-cross codes have moderate long-range connectivity and open boundary conditions, and thus are appealing candidates for near-term implementations in several qubit platforms, such as neutral atoms and trapped ions, with a moderate number of qubit movements or static long-range interactions \cite{lucas,pecorari2024highratequantumldpccodes,PhysRevLett.133.180601,bergonzoni2026faulttolerantquantumcomputationstatic}. Second, our protocol requires identifying a number of equivalent, disjoint logical operator representatives equal to the code distance, a condition readily met in La-cross codes via surface-code-inspired simple geometric arguments, as we detail below. Moreover, La-cross codes are hypergraph product codes \cite{Tillich_2014,Kovalev_2013} with logical operators aligned along single rows or columns of the main data qubit lattice. While not strictly required for the protocol's operation, this alignment is highly beneficial for near-term implementations as it reduces the overhead of parallel qubit routing, e.g., atom movements. However, the proposed logical gadget remains fully general and is not limited to La-cross codes.
Below, we discuss the necessary and sufficient conditions that a quantum LDPC code must satisfy to enable logical computation via our logical gadget, leaving the identification of specific code families and the provision of working examples beyond La-cross codes to future work.

The La-cross code family is defined as follows \cite{pecorari2024highratequantumldpccodes}: Let $\mathcal{C}=[n,k,d]$ be a cyclic classical code that encodes $k$ logical bits in $n$ physical bits with Hamming distance $d$. We denote by $r$ the number of its checks and $H\in\mathbb{F}_2^{r\times n}$ its circulant rectangular parity-check matrix, so that $k=n-\text{rank}(H)$. Let $\mathcal{C}^T$ be the transposed code of $\mathcal{C}$ with parity-check matrix $H^T\in\mathbb{F}_2^{n\times r}$, then $\mathcal{C}^T\equiv[n^T,k^T,d^T]$ with $n^T=r$, $k^T=n^T-\text{rank}(H^T)$, and $d^T$ the corresponding Hamming distance. Using the hypergraph product (HGP) \cite{Tillich_2014}, we can construct a $[[N,K,D]]$ La-cross code that has a quantum parity-check matrix
\begin{align}
\label{eq:hgp_matrix}
\begin{split}
H_Q &= \left(\begin{array}{c|c}
        0  & H_X \\
        H_Z & 0
    \end{array}\right) \\
    &= \left(
        \begin{array}{cc|cc}
        0 & 0 &  H\otimes\mathbb{I}_{n} & \mathbb{I}_{r}\otimes H^T \\
         \mathbb{I}_{n}\otimes H & H^T\otimes\mathbb{I}_{r} & 0 & 0 
        \end{array}
    \right).
\end{split}
\end{align}
This construction ensures that all the stabilizers commute, i.e. $H_XH_Z^T=0$.
If the cyclic classical code $\mathcal{C}$ is full-rank, i.e. if $r=n-k$, the resulting La-cross code inherits open boundary conditions and its quantum parameters read $N=n^2+(n-k)^2$, $K=k^2$, $D=d$ \cite{pecorari2024highratequantumldpccodes}. 

It is often convenient to describe classical cyclic codes of $n$ bits in terms of polynomials, say $h(x)$, having binary coefficients and dividing $x^n+1$ modulo $2$, i.e. $h(x)\in\mathbb{F}_2[x]/(x^n+1)$. Then, the classical cyclic code $\mathcal{C}=[n,k,d]$ that generates a La-cross code must have a circulant rectangular parity-check matrix $H=\text{circ}(1,1,0,\dots,0,1,0,\dots,0)\in\mathbb{F}_2^{(n-k)\times n}$ with $H_{00}=H_{01}=H_{0k}=1$, which corresponds to a generating polynomial $h(x)=1+x+x^k$. This classical code can be regarded as a higher-rate generalization of the classical repetition code, $h_{rep}(x)=1+x$. We also note that tensoring two copies of a classical code with weight-$3$ checks via HGP produces stabilizers of weight-$6$ in the resulting La-cross code [see Fig.~\ref{fig:fig1}(a)].

\begin{figure}[t!]
    \centering
    \includegraphics[width=\linewidth]{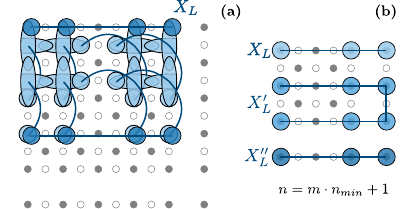}
    \caption{(a) Logical operators in a La-cross code can be translated along the main lattice via multiplication by a subset of stabilizers. For $k=2$, an example is illustrated: Dark blue strings are two equivalent representations of the same logical operator $X_L$, while in light blue we show the patter of $X$-stabilizers that translates the logical operator in the first row three rows downward. (b) For $k=2$ codes, for some lattice sizes (i.e., $n=m\cdot n_{min}+1$ for some $m\in\mathbb{N}$) and open boundary conditions, some logical operators can be longer than the code distance $d$. Here we show the case of a $d=2$ $k=2$ La-cross code which also admits $d+1=3$-long logical operators. Crucially, it is always possible to find as many non-overlapping equivalent representatives ($X_L$, $X_L'$, $X_L''$ logical strings colored with different gradients of blue) via multiplication by a suitable subset of stabilizers. Notably, some of these equivalent representations can have length $2d$ (e.g, $X_L'$ in the example shown). This observation is key to enable the fault tolerance of our quantum computing gadget.
    }
    \label{fig:fig2}
\end{figure}

The logical operators of a $[[N,K,d]]$ QEC code are Pauli strings in the $N$-qubit Pauli group, $L\in\mathcal{P}_N$, which commute with all stabilizers without being a product of stabilizers. For CSS codes, 
\begin{equation}
    L=\left(\begin{array}{c|c}
        0  & L_X \\
        L_Z & 0
    \end{array}\right)\;.
\end{equation}
It follows that $L_Z\in\ker{(H_X)}$, but $L_Z\notin\text{im}{(H_Z^T)}$, and $L_Z\in\ker{(H_Z)}$, but $L_X\notin\text{im}{(H_X^T)}$. It is then easy to verify that the shortest logical operators of any La-cross code are horizontal and vertical Pauli strings that always align along a single row or column of the lattice [see Fig.~\ref{fig:fig1}(a) for an example] \cite{pecorari2024highratequantumldpccodes}.

Consider a $[[N,K,d]]$ La-cross code with open boundaries and let $X_L$ be one of its $K=k^2$ logical operators of $X$-type. To perform addressable logical operations on a specific logical qubit encoded in a La-cross code in a fault-tolerant manner, our protocol requires finding the shortest $d$ non-overlapping equivalent representatives of $X_L$, that is, the shortest $X$-strings that are equivalent to $X_L$ up to multiplication by some $X$-stabilizers. 

Without loss of generality, we assume that $X_L$ has support over $d$ physical qubits in the first row of the lattice. To translate a logical operator defined on the first row of the main lattice downward while preserving its length and pattern, we proceed as follows -- the generalization to logical operators that do not have support on the first row is straightforward and follows from similar considerations. For the sake of clarity, we restrict the discussion to the case of $k=2$ La-cross codes and refer to the Methods section for a general discussion.
First, we identify the X-stabilizers that overlap on one qubit with the logical operator $X_L$. Then, we multiply $X_L$ by the two subsequent rows of $X$-stabilizers with the pattern just identified [for an example of such stabilizer pattern, see the light blue operators in Fig.~\ref{fig:fig2}(a)]. This effectively translates the $X_L$ logical operator three rows downward in the main lattice. From this new equivalent representation of $X_L$, we repeat the same process several times and stop when the boundary of the lattice is reached. In this way, $d'<d$ equivalent $d$-long representatives are found. Each of these strings, say $X_L$, $X_L'$, $X_L''$, ..., is then multiplied by a single row of $X$-stabilizers which are just below each string and overlap on one qubit with it. Such a row of stabilizers always exists except if the logical Pauli string we are considering is close to the boundary of the lattice. In this way, we find the remaining $d-d'$ non-overlapping equivalent representatives of the original $X_L$ logical operator. We observe that these last $d-d'$ representatives are $2d$-long $X$-strings with the same pattern as the original $X_L$, but spanning across two rows of the main lattice [see, for example, Fig.~\ref{fig:fig2}(b), where different equivalent representations of the same logical operator are represented as strings colored with different gradients of blue].

For higher-rate, i.e. $k\geq4$, La-cross codes and certain values of $n$, we find that some of the shortest, non-overlapping, and equivalent representatives of a logical operator can span across $\ell$ rows, where $\ell$ is upper bounded by $k-1$. We comment more in details on those cases in the Methods section.

As an interesting observation, we find that, when open boundary conditions are enforced, the stabilizer pattern of La-cross codes breaks the standard notion of translational invariance across the lattice. As a consequence, a distance-$d$ La-cross code has at least one pair of $d$-long $(X_L,Z_L)$ logical operators, but it also admits longer logical operators. This can be easily understood as follows. Consider a $k=2$ La-cross code, which has stabilizers of $Z$-type (the discussion is symmetric for $X$) acting on three qubits aligned in a row, say, $q_0q_1q_2$. Therefore, $\overline{X}_0=X_0X_1$, $\overline{X}_1=X_0X_2$, and $\overline{X}_2=X_1X_2=\overline{X}_0\overline{X}_1$ all commute with the stabilizers. Additionally, $k=2$ La-cross codes enjoy translational invariance over $n_{min}=3$ lattice sites. The logical operators can be easily identified by repeating the (not all linearly independent) generating patterns $\overline{X}_0,\overline{X}_1,\overline{X}_2$ along one row of the $n\times n$ principal lattice. When $n=m\cdot n_{min}+1$, for some $m\in\mathbb{N}$, half of the four logical operators will then have length $d$ and half will have length $d+1$.  
Nevertheless, it is always possible to find as many non-overlapping equivalent representatives as the length of the longest logical operator with the iterative procedures explained above [see Fig.~\ref{fig:fig2}(b)]. Crucially, this can be leveraged to engineer a fault-tolerant protocol that preserves the circuit-level distance of each distinct logical operator, even of those that are longer than the actual distance of the code. Preserving the circuit-level distance of each individual logical operator -- especially those that are longer than $d$ -- is highly relevant, e.g., for the compilation of algorithms that only use a subset of the logical qubits encoded in a quantum LDPC code. This would allow one to select only the longest logical operators and hence improve robustness against errors.  

Finally, we observe that similar considerations can be generalized to higher-rate, i.e. $k>2$, La-cross codes with open boundaries that break translational invariance on a larger scale. For example, by inspection, we find that $n_{min}=7$ for $k=3$ and $n_{min}=15$ for $k=4$. In general, we find the following upper-bound: $n_{min}\leq2^k-1$, which scales exponentially in $k$. It follows that codes with large $k$ more often admit logical operators which are longer than $d$ physical qubits.

\subsection{Auxiliary Bacon-Shor code}
The quantum computing gadget that we design to enable addressable quantum computation with La-cross quantum LDPC codes leverages gate teleportation via an auxiliary code coupled transversally to the target logical qubit. We first present this construction using the Bacon-Shor (BS) code as the auxiliary code, since its simple gauge structure makes the coupling mechanism, and its correctness proof, most transparent. The BS gadget is additionally attractive for near-term implementations, as, as we will show, it has a substantially lower gate count than a comparable auxiliary subspace code. We show below, however, that because the BS code is a subsystem rather than a subspace code, the resulting gadget admits only a finite pseudo-threshold rather than a genuine asymptotic threshold. Below, in a dedicated section, we present a scalable extension of the same construction using the unrotated surface code as the auxiliary code instead. This recovers a genuine threshold at the cost of an additional sub-lattice completion absent in the BS case, while leaving the main-lattice coupling, and the mechanism of Fig.~\ref{fig:fig1}(d)-(g), unchanged.

The Bacon–Shor (BS) code \cite{baconPhysRevA.73.012340} $[[N_{BS},1,d_{BS}]]$ is a subsystem stabilizer code that encodes one logical qubit into a $n\times m$ lattice of physical qubits. The number of independent stabilizers is strictly smaller than $N_{BS}-1$, hence certain degrees of freedom do not store logical information and can be freely modified without affecting the encoded state. Let $X_{i,j}^{\text{BS}}$ ($Z_{i,j}^{\text{BS}}$) be a $X$ ($Z$) operator acting on the qubit in the $i$th row and $j$th column of the BS lattice. The BS code is defined by $n + m -2$ stabilizers 
\begin{eqnarray}
    S_{z,j}^{\text{BS}} = \prod_{i=1}^nZ_{i,j}^{\text{BS}} Z_{i,j+1}^{\text{BS}},\qquad
    S_{x,i}^{\text{BS}} = \prod_{j=1}^mX_{i,j}^{\text{BS}} X_{i+1,j}^{\text{BS}}
\end{eqnarray}
which span two entire rows or columns and are then highly non-local [see Fig.~\ref{fig:fig1}(b)]. The remaining degrees of freedom can be used to define local gauge operators [see Fig.~\ref{fig:fig1}(c)],
\begin{eqnarray}
    G_{z,i,j}^{\text{BS}} = Z_{i,j}^{\text{BS}} Z_{i,j+1}^{\text{BS}},\qquad
    G_{x,i,j}^{\text{BS}} = X_{i,j}^{\text{BS}} X_{i+1,j}^{\text{BS}},
\end{eqnarray} 
such that the original non-local stabilizers can be recovered as the product of weight-$2$ gauge operators,
\begin{eqnarray}
    S_{z,i}^{\text{BS}} =\prod_{i=1}^n G_{z,i,j}^{\text{BS}},\qquad
    S_{x,j}^{\text{BS}} = \prod_{j=1}^m G_{x,i,j}^{\text{BS}}\;.
\end{eqnarray}
Thus, the BS code has $n(n-1)+m(m-1)$ non-mutually commuting gauge operators. A gauge is fixed by promoting a maximal commuting subset of gauge operators to stabilizers. For example, in the so-called \emph{$Z$-gauge} all $Z$-type gauge operators are chosen as $Z$-stabilizers, $^ZS_{z,i,j}^{\text{BS}} = G_{z,i,j}^{\text{BS}}$, while in the \emph{$X$-gauge} all the $X$-type gauge operators are chosen as $X$-stabilizers, $^XS_{x,i,j}^{\text{BS}} = G_{x,i,j}^{\text{BS}}$. 
Additionally, we observe that the $X_L$ ($Z_L$) logical operator of the BS code is an $X$ string acting on a single row (column) of the lattice [see Fig.~\ref{fig:fig1}(c)]. 

Finally, we note that there exists another possible gauge fixing choice for the BS code, which we call the \emph{surface code gauge}, that effectively morphs the BS code into a rotated surface code \cite{Li2018}. This is convenient as it allows one to use the full surface code toolbox and then morph the code back to a BS code. In the following, we will use this possibility to implement a S gate directly in the BS code without extra qubit overhead.

\subsection{Addressable Hadamard rotation}
We start by presenting the protocol to perform a Hadamard (H) gate, i.e., a logical single-qubit rotation $\mathcal{R}_y(\pi/2)$, on a selected logical qubit encoded in the La-cross code [Fig.~\ref{fig:fig1}(d)]. 

We consider a $d$-long $X_L$ logical operator of the La-cross code with support on the first row of the lattice, together with its $d$ non-overlapping, shortest, and equivalent representatives, as discussed before. The collection of these equivalent logical operators effectively realizes a partition of the La-cross code lattice. In this way, the only physical qubits involved in the operations are those that uniquely correspond to the single La-cross logical qubit we aim to address. 

The protocol for performing a Hadamard rotation is as follows. (\emph{i}) We first consider a $d\times d$ square BS code and initialize it in its $|+\rangle^{\text{BS}}$ logical state. For the state preparation and all the controlled operations to be fault-tolerant, we use the $Z$-gauge for the BS code.
Next, (\emph{ii}) we apply a logical transversal CNOT (tCNOT) gate controlled by the rows of the BS code and targeting all the equivalent representatives of the selected $X_L$ logical operator of the La-cross code, that is
\begin{equation*}
\ket{+}^\text{BS}\ket{\psi}^\text{qLDPC}\xrightarrow{\text{tCNOT}}\frac{1}{\sqrt{2}}\left(\ket{0}^\text{BS}+\ket{1}^\text{BS}X\right)\ket{\psi}^\text{qLDPC}.
\end{equation*}
Then, (\emph{iii}) we apply a logical transversal CZ (tCZ) gate controlled by the rows of the BS code and that targets all the equivalent representatives of the selected $Z_L$ logical operator of the La-cross code (that is, the $Z_L$ logical operators corresponding to the initial $X_L$ logical operator),
\begin{align*}
    \begin{split}
        &\xrightarrow{\text{tCZ}}\frac{1}{\sqrt{2}}\left(\ket{0}^\text{BS}+\ket{1}^\text{BS}ZX\right)\ket{\psi}^\text{qLDPC}\\
        &=\frac{\ket{+}^\text{BS}}{\sqrt{2}}\frac{1+iY}{\sqrt{2}}\ket{\psi}^\text{qLDPC}+\frac{\ket{-}^\text{BS}}{\sqrt{2}}\frac{1-iY}{\sqrt{2}}\ket{\psi}^\text{qLDPC}.
    \end{split}
\end{align*}
We show in Fig.~\ref{fig:fig1}(e) a possible implementation scheme for these two transversal controlled operations by using either qubit shuttling or long range couplers.
We also note that a round of QEC stabilizer readout has to be inserted on both codes between the two controlled logical operations to ensure fault tolerance by preventing errors from
propagating to multiple physical qubits without being detected. 
Finally, (\emph{iv}) we measure the BS code in the $X$-basis. Thus, if the BS code is measured in the $|+\rangle^{\text{BS}}$ state, the protocol correctly implements a $\mathcal{R}_y(\pi/2)$ rotation of the selected La-cross logical qubit. Instead, if the BS code is in state $\ket{-}^{\text{BS}}$ before the measurement, a conditional logical $Y$ correction must be applied to the targeted La-cross logical qubit. Notably, such a correction can be tracked classically throughout the computation when the remaining circuit consists only of Clifford gates, and when non-Clifford states are prepared offline and injected via teleportation protocols. 

We note that our protocol only requires measuring the $Z$-stabilizers of the BS code and never the $X$-stabilizers, which we then only measure during the final $X$-measurement. This is due to the fact that $Z$-errors in the BS code can never spread to the targeted logical La-cross qubit via the two controlled operations, which reduces the gate cost of the protocol. 
We also observe that the same protocol can be used to implement the same gate with opposite angle, that is, a $R_y(-\pi/2)$ rotation, either by reversing the order of the controlled operations or by classically controlling the $Y$-correction with the $\ket{+}^{\text{BS}}$ logical state of the BS code.

Finally, we anticipate that this same protocol structure with steps (\emph{i})-(\emph{iv}) is unchanged for the scalable surface code extension that we present below, where only the auxiliary code and transversal coupling differ.

In the next section, we discuss the circuit-level fault tolerance of our Hadamard rotation protocol and show QEC numerical simulations to assess its logical error-correction performance.

\subsection{Circuit-level fault tolerance of the BS gadget}
We now show that the transversal nature of the controlled operations between the BS code and the selected qubit partition of the La-cross code ensures circuit-level fault tolerance, i.e., that a single physical fault cannot spread to a spatially uncorrectable error, that the operations correctly implement the intended logical gate, and that stabilizers of both codes propagate correctly through them. 

First, circuit-level fault tolerance straightforwardly derives from the transversality of the protocol, which individually couples each of the $d$ rows of the BS code with each of the $d$ different representatives of the selected logical operator in the La-cross code [see also Fig.~\ref{fig:fig1}(e)]. Crucially, the identified equivalent representatives of the same logical operator must have non-overlapping support for the protocol to be fault-tolerant. The vulnerabilities introduced by using overlapping equivalent logical representatives are detailed in the Methods section. We also observe that the longer, namely $2d$-long (or, generally, $\ell d$-long), representatives can be fault-tolerantly controlled with a $d$-long row of the BS code provided that the operation is realized in two ($\ell$) steps, one for each of the two ($\ell$) $d$-long rows of the logical representative. That is because an $X$ error on one of the control qubits of the BS code can only propagate through the CNOT gates into vertical pairs ($\ell$-tuple) of $X$ errors on the La-cross code. Since the logical $X_L$ operators are all horizontal as they all and only lay in the main data qubit lattice of the La-cross code, these vertical $X$ errors do not align with the logical operator, and hence do not degrade the circuit-level distance of the code, thus ensuring fault tolerance. The reasoning is similar for CZ gates.

We now analytically prove that the protocol implements the desired logical operation. To do so, it is sufficient to show that the tCNOT operation between the Bacon-Shor code and the partitioning of the La-cross code lattice is well defined, that is, it implements the intended logical CNOT and preserves the joint codespace up to products of stabilizers. That is, formally, $\text{tCNOT} \cdot S_i\cdot \text{tCNOT}= S_i\prod_{j\neq i}S_j,\,\forall i,$ where $S_{i} $ is a stabilizer of either code. Symmetrical considerations hold for the transversal CZ, as it only differs by a change of basis.

We start by showing that the entangling operations is a logical CNOT. The logical state $\ket{0}^{\text{BS}}$ of the BS code is a superposition of all states with an even number of rows in the state $\ket{\bar{1}}=\ket {1\cdots1}$. In contrast, its logical state $\ket{1}^{\text{BS}}$ is a superposition of all states with an odd number of such rows. Therefore, when the BS code is in $\ket{0}^{\text{BS}}$ ($\ket{1}^{\text{BS}}$), the $X_L$ logical operator of the targeted La-cross logical qubit is applied an even (odd) number of times, thereby acting as the identity (single Pauli $X$) on the targeted qubit. Thus, this proves that such operation correctly implements a logical tCNOT on the selected La-cross logical qubit controlled by the BS code.

We now show that this operation is correctly implemented up to products of stabilizers of either code, proving that the codespaces of both Bacon-Shor and La-cross codes are preserved. Let us consider how the stabilizers of each code propagate through the tCNOT. First, we observe that, due to transversality, the $Z$-stabilizers of the BS code and the $X$-stabilizers of the La-cross code commute trivially through the tCNOT, and are therefore left unchanged. Instead, a $Z$-stabilizer of the La-cross code propagates through the tCNOT into a product of $Z$ Pauli operators on the BS code.  
Furthermore, since the $X_L$ logical operator and the $Z$-stabilizers of the La-cross code commute by definition, they always overlap on an even number of qubits in the main lattice. Consequently, after conjugation by tCNOT, the $Z$-stabilizers of the La-cross code transform as
\begin{equation}
     \text{tCNOT} \cdot S_{z,i}^{\text{qLDPC}} \cdot \text{tCNOT} = S_{z,i}^{\text{qLDPC}}\cdot Z_{k_i,j_1}^{\text{BS}} \cdot Z_{k_i,j_2}^{\text{BS}}
     \label{eq:ZLDPC_CNOT}
\end{equation}
where $S_{z,i}^{\text{qLDPC}}$ denotes the $i$-th $Z$-stabilizer of the La-cross code and  $Z_{k_i,j}^{\text{BS}}$ is the Pauli $Z$ operator acting on the data qubit located in the $k_i$-th row and $j$-th column of the BS code. In this protocol, we require fixing the $Z$-gauge of the BS code, where we promote to stabilizers the two-body gauge operators $^ZS_{z,i,j}^\text{BS}:=Z_{i,j}^\text{BS}\cdot Z_{i,j+1}^{\text{BS}}$. Therefore, upon fixing the $Z$ gauge, the product $ Z_{k_i,j_1}^{\text{BS}} \cdot Z_{k_i,j_2}^{\text{BS}}$ in Eq.~(\ref{eq:ZLDPC_CNOT}) can be expressed as a product of BS $Z$-stabilizers as $$ Z_{k_i,j_1}^{\text{BS}} \cdot Z_{k_i,j_2}^{\text{BS}}=\prod_{j'=j_1}^{j_2-1} \ ^ZS_{z,i,j'}^\text{BS},$$
proving that the $Z$ stabilizers of the La-cross commute through the tCNOT up to a product of stabilizers of the BS code.
Let us now consider the $X$ stabilizers of the BS code: since these act on two consecutive rows of data qubits, under conjugation by tCNOT they acquire support on two consecutive, non-overlapping representatives of the same $X_L$ logical operator of the La-cross code. Since  different representatives of a La-cross logical operator differ by a product of $X$ stabilizers, it follows that $$\text{tCNOT} \cdot S_{x,i}^{\text{BS}} \cdot \text{tCNOT} = S_{x,i}^{\text{BS}}\cdot \prod_j S_{x,j}^\text{qLDPC},$$ where $S_{x,i}^\text{BS}$ denotes the $i$-th $X$-type stabilizer of the BS code and $S_{x,j}^\text{qLDPC}$ are $X$-stabilizers of the La-cross code. Thus, also the $X$ stabilizers of the Bacon-Shor code commute through the tCNOT up to a product of stabilizers of the La-cross code. This concludes the proof that our protocol implements the correct logical operation, while preserving the codespaces of both BS and La-cross codes.

Finally, we note that the BS gadget implements the desired operations on a single La-cross logical qubit, while leaving the other logical qubits unaffected. That is, similarly to the logical Pauli operators, which only affect their corresponding logical qubit, their controlled version used in the BS gadget does not affect the logical state of the surrounding logical qubits.

\subsection{Scalable fault tolerance and surface code extension}
\label{sec:sc}

The circuit-level fault tolerance established above guarantees that a single physical fault remains a single, spatially correctable error at any fixed code distance. It does not, however, guarantee that the logical error rate continues to improve as the distance grows. The BS code is known to possess no genuine asymptotic threshold but only a finite pseudo-threshold \cite{gidney2023baconthreshold}, an issue that is inherited by our logical gadget even though it touches the BS code for only a constant number of rounds. This follows from two compounding effects, both rooted in the fact that the BS code is a subsystem rather than a subspace code. First, in the $Z$-gauge, each row's $Z$-type value is never independently corrected during the circuit. Instead, it is read out only at the end, as an unprotected length-$d$ classical parity of $d$ physical measurement outcomes, whose own error probability grows with $d$ rather than decreasing, and the logical value is decided by majority vote across $d$ such degrading row parities. Second, the $X$-stabilizers of the BS code are never measured during the circuit but are reconstructed at the end from the final $X$-measurement of all data qubits. Such reconstruction similarly combines a growing number of equally noisy outcomes as $d$ increases. Both effects worsen with code distance, and together they imply that there exists a finite distance $d^*$, beyond which the BS gadget's logical error probability starts to increase with $d$ rather than decreasing, allowing for only a finite-size pseudo-threshold.

\begin{figure}[t!]
    \centering
    \includegraphics[width=\linewidth]{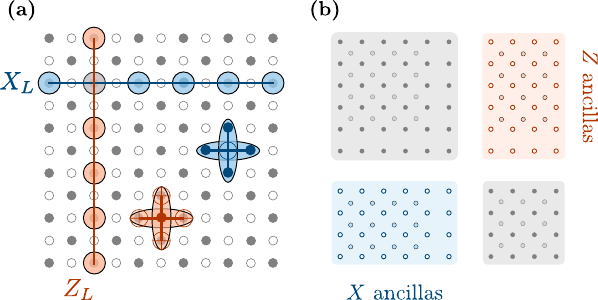}
    \caption{(a) Unrotated surface code, two stabilizer operators and the two logical $X_L$ and $Z_L$ operators are drawn in blue and red, respectively. (b) Both La-cross and unrotated surface code are examples of hypergraph product codes. The panel shows how they can be laid down in data and ancilla blocks, and interleaved to facilitate the transversal coupling. }
    \label{fig:surf-code}
\end{figure}

Recovering a genuine asymptotic threshold requires replacing the BS code with a full two-dimensional stabilizer code. We show in this section that this can be done while preserving the entire gadget structure introduced above -- notably the main-lattice coupling described above is unchanged -- by using a $d\times d$ unrotated surface code [Fig.~\ref{fig:surf-code}(a)] as the auxiliary code, with $d$ the quantum LDPC code distance. Unlike the BS code, the unrotated surface code is itself a hypergraph-product code [Fig.~\ref{fig:surf-code}(b)], with stabilizers supported on both a main lattice and a secondary sublattice. As a consequence, the main-lattice coupling alone  (sufficient for the BS gadget, whose gauge operators are one-dimensional and involve no sublattice) no longer maps surface-code and La-cross stabilizers to products of stabilizers. A sub-lattice coupling between the two codes' sublattices, with no analogue in the BS gadget, is required to fix this issue. Notably, we find that this completion can be solved with coupling degree independent of the code distance, so that the resulting surface-code gadget retains the same constant-depth structure as the BS gadget while restoring a genuine asymptotic threshold and preserving the $\mathcal{O}(1)$ time overhead of our protocol.

\begin{figure}[t!]
    \centering
    \includegraphics[width=\linewidth]{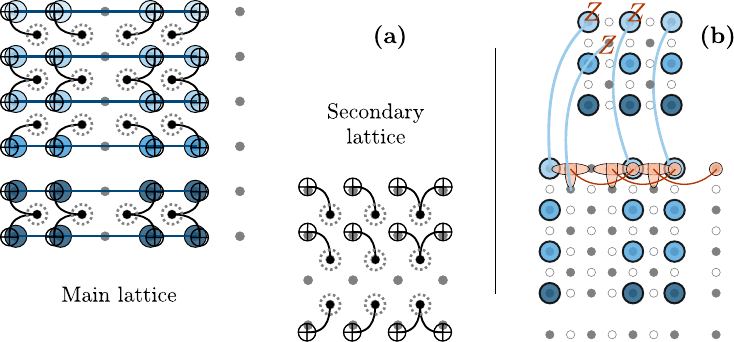}
    \caption{(a) Transversal CNOT coupling of La-cross code partition and an unrotated surface code in both main (left) and secondary (right) lattice. Surface code's qubits are gray dashed circles. (b) The sub-lattice coupling is to guarantee that the codes' stabilizer are mapped into products of stabilizer upon transversal coupling. The example here is for the propagation of the La-cross code's $Z$-stabilizers (in red). The same coupling also ensures the correct propagation of the surface code's $X$-stabilizers.}
    \label{fig:sc-coupling}
\end{figure}

The main-lattice coupling is identical to that of the BS gadget, with a transversal CNOT (CZ) which pairs each row of the $d\times d$ surface code with one of the $d$ non-overlapping representatives of the addressed $X_L$ ($Z_L$) logical operator of the La-cross code. Unlike the BS code's non-local stabilizers, the surface code's $X$-stabilizers are local, each acting on four (in the bulk) or three (at the boundaries) neighboring data qubits. Under conjugation by the tCNOT on the main lattice only, the $Z$-stabilizers of the La-cross codes are mapped onto the La-cross code as
\begin{align*}
    \begin{split}
        \text{tCNOT}^\text{main} &\cdot S_{z,i}^{\text{qLDPC}} \cdot \text{tCNOT}^\text{main} \\
        &= S_{z,i}^{\text{qLDPC}}\cdot Z_{k_i,j_1}^\text{SC, main}\cdot Z_{k_i,j_2}^\text{SC, main}
    \end{split}
\end{align*}
where $Z_{k_i,j}^\text{SC, main}$ is the Pauli $Z$ operator acting on the data qubits located at row $k_i$ and column $j$ of the surface code (SC) main lattice. The SC's stabilizers act on both main and secondary lattice. For instance, their bulk surface code stabilizers can be defined as follows,
\begin{equation}
    S^{\text{SC}}_{z,i,j}= Z_{i,j}^\text{SC, main} \cdot Z_{i,j+1}^\text{SC, main}\cdot Z_{i,j}^\text{SC, sec.} \cdot Z_{i+1,j}^\text{SC, sec.}.
\end{equation}
The sub-lattice coupling between both data qubit blocks serves as stabilizer completion [Fig.~\ref{fig:sc-coupling}(a)] to guarantee that stabilizers correctly map into stabilizers via tCNOT. Formally, in the bulk we have
\begin{align*}
    \begin{split}
        &\text{tCNOT}\cdot S_{z,i}^{\text{qLDPC}} \cdot \text{tCNOT} \\
        &=S_{z,i}^{\text{qLDPC}}\cdot \prod_{m=j_1}^{j_2-1}S^{\text{SC}}_{z,k_i,m}
        \\
        &=S_{z,i}^{\text{qLDPC}}\cdot Z_{k_i,j_1}^\text{SC, main}\cdot Z_{k_i,j_2}^\text{SC, main} \cdot \prod_{m=j_1}^{j_2-1}Z_{k_i,m}^\text{SC, sec.}\cdot Z_{k_i+1,m}^\text{SC, sec.}.
    \end{split}
\end{align*}
The above equation can be satisfied by imposing $S_{z,i}^{\text{qLDPC}}$'s data qubits on the secondary lattice to couple to surface code 's data qubits on the secondary lattice located at rows $k_i$ and $k_i+1$ and columns from $j_1$ to $j_2-1$. For visual intuition, Fig.~\ref{fig:sc-coupling}(b) shows an example for boundary $Z$-stabilizers. It is important to observe that the coupling introduced in the secondary lattice does not affect the CNOT's logical action, since hypergraph product codes with open boundary conditions -- such as La-cross and unrotated surface codes -- have  the shortest representative of the logical operators that \emph{only} live in the main lattice of data qubits. Additionally, even though the coupling is not strictly transversal on the secondary lattice, the code distance remains unaffected because of the such property.

Symmetric expressions govern the flow of surface code's $X$-stabilizers.
Since $j_2$ is at most $j_1+k$, the coupling connectivity  on the secondary lattice scales at most linearly with $k$ and, more importantly, is code distance independent, so the resulting SC gadget retains the constant-depth structure of the BS gadget [Fig.~\ref{fig:sc-coupling}(a)].

Crucially, this construction resolves the limitation identified above without sacrificing any of the properties that make the BS gadget attractive. The surface code is a stabilizer code with a well-defined asymptotic threshold, which is now inherited by our logical gadget. Moreover, the constant-depth of the transversal coupling preserves such threshold ensuring only a constant number of QEC rounds is necessary and enables the $\mathcal{O}(1)$ time overhead of our construction. This solely comes at the price of a modest increase in gate count relative to the BS gadget, due to the different stabilizer readout and the sub-lattice coupling itself.

\subsection{Addressable single qubit rotations}
Similarly to the Hadamard gate, also the S gate, i.e. a logical single-qubit rotation $\mathcal{R}_z(\pi/2)$, can be implemented on any single La-cross logical qubit via gate teleportation. 

We start by initializing the $d\times d$  square BS code in state $|S\rangle\equiv S|+\rangle^{\text{BS}}$. Then, a logical tCZ gate between the BS code and the targeted La-cross logical qubit is applied, and finally the BS code is measured in the $X$-basis. A $Z$-gate correction is applied, depending on the outcome of the $X$-measurement. Therefore, this protocol [the circuit is depicted in Fig.~\ref{fig:fig1}(f)] correctly implements an S gate on the targeted La-cross logical qubit. 
We also observe that a single-qubit rotation $\mathcal{R}_x(\pi/2)$ can be implemented in a very similar manner, using a logical tCNOT gate instead of the tCZ in the circuit described above. 

The protocol we have just described exploits the fact that an $\ket{S}$ state can be fault-tolerantly prepared in the BS code by initializing it in the $\ket{+}_L$ of the surface code gauge~\cite{Li2018},
where fault-tolerant methods for implementing an S gate are known. In particular, a fold-transversal S gate can be fault-tolerantly implemented halfway through the syndrome extraction circuit leveraging the equivalence between rotated and unrotated surface codes at half-cycle \cite{chen2024transversallogicalcliffordgates}. After performing this operation, we morph the BS code back to the $Z$-gauge in order to fault-tolerantly apply the subsequent logical controlled operation.

Instead, with the surface code gadget, the $S$ gate protocol simplifies further, since the unrotated surface code naturally supports the logical $S$ gate transversally via folding \cite{Moussa_2016}.

\subsection{Addressable two- and multi-qubit gates}
The protocol for implementing logical multi-qubit entangling gates is a generalization of the scheme discussed above for the Hadamard rotation. Notably, the targeted logical qubits can now belong to either the same La-cross code or to two distinct codes.

We start by presenting how to realize a logical two-qubit entangling gate.
The scheme is as follows: The $d\times d$ auxiliary code is first initialized to its logical $|+\rangle^{\text{BS/SC}}$ state and then entangled via two tCNOT gates to each La-cross logical qubit in the same way as in the previous sections. Then, a tCZ gate is applied between the auxiliary code and the first La-cross logical qubit. Finally, the auxiliary is measured in the $X$-basis. Conditionally on the auxiliary measurement outcome, a $Y_1X_2$ correction is applied on the two targeted logical qubits. This protocol then correctly realizes an entangling logical $e^{i\frac{\pi}{4}Y_1 X_2}$ gate [see also Fig.~\ref{fig:fig1}(g)]. 

In the same spirit, we observe that, if a tCZ is applied in place of the first tCNOT, a $e^{i\frac{\pi}{4}Y_1 Z_2}$ gate is realized. Additionally, using the S-gate protocol described above, our scheme allows one to perform any two-qubit gate of the form $e^{i\frac{\pi}{4} X_1 X_2}$, $e^{i\frac{\pi}{4} Z_1 Z_2}$, $e^{i\frac{\pi}{4} Y_1 Y_2}$ and $e^{i\frac{\pi}{4} X_1 Z_2}$. This can then be straightforwardly extended to more than two logical qubits, enabling the native implementation of any multi-qubit logical gate of the form $e^{i\frac{\pi}{4} P}$, where $P$ is any Pauli string.

\begin{figure}
    \centering
    \includegraphics[width=\linewidth]{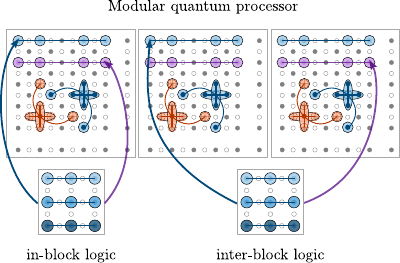}
    \caption{To demonstrate the versatility of the proposed logical gadget, we envision a modular quantum processor wherein La-cross code units are interleaved with shuttling auxiliary codes. This architecture facilitates the generation of robust entanglement both within individual physical code blocks and across spatially separated logical qubits.
    }
    \label{fig:entangling}
\end{figure}

All these observations suggest that, since the Clifford gate set enabled by our logical gadget is not minimal, efficient circuit-recompilation strategies may exploit this richer native gate set to reduce the overall gate count for quantum algorithms applications. In addition, the execution of these logical entangling gates is agnostic to the underlying block connectivity. Because the interacting logical strings can belong to the same code block or to different ones, our logical gadget demonstrates significant architectural flexibility, and can generate both intra-block and inter-block entanglement across spatially separated code units, as illustrated in Fig.~\ref{fig:entangling}.

We also note that, in contrast to the measurement-based approach~\cite{Cohen2022,xu2023constantoverhead,Cross2024,Williamson2024,Cowtan2025,Ide2025,Swaroop2025}, which requires $\mathcal{O}(d)$ rounds of QEC to perform a logical operation, our scheme operates in a time that is independent of the code distance. In fact, the total duration of a logical gate of the form $ e^{i\frac{\pi}{4} P}$ depends solely on the number of controlled operations needed. In turn, the number of necessary controlled operations corresponds to the number of non-trivial Pauli operators in the Pauli string $P$ defined above, and counting each $Y$ operator twice, since it decomposes into $X$ and $Z$. Moreover, several optimizations can reduce this number by merging multiple controlled logical operations into a single one. For example, (\emph{i}) if two logical operators $X_{L_1},X_{L_2}$ act on the same row, the corresponding controlled operation can be applied directly to $X_{L_1}\cdot X_{L_2}$, that is, to the qubits in the support of the product of both operators. This is possible if the product of two $d$-long logical operators is still a $d$-long logical Pauli string, which is exactly the case for those logical operators that are as long as the distance $d$ of the code. 
Alternatively, (\emph{ii}) for small-$k$ La-cross codes, if two logical operators act on different rows but have the same pattern, it is possible in some cases to find a compact (i.e., single-row) representation of their product on another row. The protocol can then be applied to this compact representation, and the physical controlled operation performed on the different representatives — where the two original rows jointly form a 
$2d$-long representative of the compact operator. Similar considerations apply to $Z$ logical operators. 

Finally, as an example of use-case, we observe that our logical gadget can be used to implement a Greenberger–Horne–Zeilinger (GHZ) state of all the logical qubits encoded in a single La-cross code. This can be achieved by preparing one logical qubit in state $|+\rangle^\text{qLDPC}$ and the others in state $|0\rangle^\text{qLDPC}$, and then performing a cascade of CNOT gates -- which are available up to single-qubit logical rotation with our protocol -- between any pair of logical qubits.

\subsection{Alternative approach: Pauli-based quantum computation}
So far, we have discussed unitary Clifford gates for digital gate-based quantum computing. However, one can adopt a different paradigm for quantum computation based on measurements of joint logical operators, which enables the implementation of any Clifford circuit. This approach, known as Pauli-based computation (PBC)~\cite{Bravyi2016}, underlies many lattice-surgery techniques for the surface code~\cite{Horsman2012} and for quantum LDPC codes~\cite{Cohen2022,xu2023constantoverhead,Cross2024,Williamson2024,Cowtan2025,Ide2025,Swaroop2025}. This mechanism enables the generation of entanglement across arbitrary subsets of qubits. Together with the standard single-qubit Clifford gates, any of these entangling operations suffices to generate the full Clifford group.

This additional framework can be used to augment the native Clifford gate set enabled by our protocol.
Using our logical gadget, it is possible to measure any Pauli operators. This is achieved by initializing the auxiliary code in the logical state $\ket{+}^\text{BS/SC}$ and applying a logical $\mathrm{CNOT}$ (resp. $\mathrm{CZ}$) gate to a target logical qubit whenever a $X_L$ (resp. $Z_L$) logical operator appears in the joint product or both gates if it is  $Y_L=iX_LZ_L$ operator. Additionally, if there is an odd number $Y_L$ logical operator involved in the Pauli product, an $S$ gate needs to be inserted after the initialization to prepare the gadget in the logical state $\ket{+}_Y$. By subsequently measuring the auxiliary code in the $X$-basis, one effectively implements a projective measurement $\frac{1\pm P}{2}$, where $P$ denotes the product of the logical operators. 

Similarly to the gate-based paradigm discussed in the previous section, our logical gadget also allows one to generate a logical GHZ state (up to Pauli corrections) between all the logical qubits encoded in a single La-cross code by first initializing all the logical qubits in $\ket{+}^\text{qLDPC}$ and then measuring the product $Z_i^L Z_{i+1}^L$ for each pair of consecutive qubits. This would require $K-1$ logical gadgets, $K$ being the number of logical qubits of the La-cross code.

\subsection{Towards addressable non-Clifford operations}
Finally, we comment on how one could complement our logical gadget with non-Clifford resources to enable universal gate-based quantum computation.

Similarly to Clifford resources, non-Clifford operations of the form \(e^{i\frac{\pi}{8}P}\), where \(P\) is the same Pauli string defined above, can be prepared in the auxiliary code and then teleported into the quantum LDPC code. 
In close analogy with the $S$-gate scheme, one could fold-transversally cultivate the $|T\rangle$ state in the surface code~\cite{sahay2025foldtransversalsurfacecodecultivation} and, in case, then morph it into the BS code by measuring its $Z$-gauge stabilizers. As an alternative approach, we could directly cultivate the $T$ state in the folded BS code, which we let as an open future perspective. 

\subsection{Numerical simulations}

\begin{figure}[t!]
    \centering
    \includegraphics[width=\linewidth]{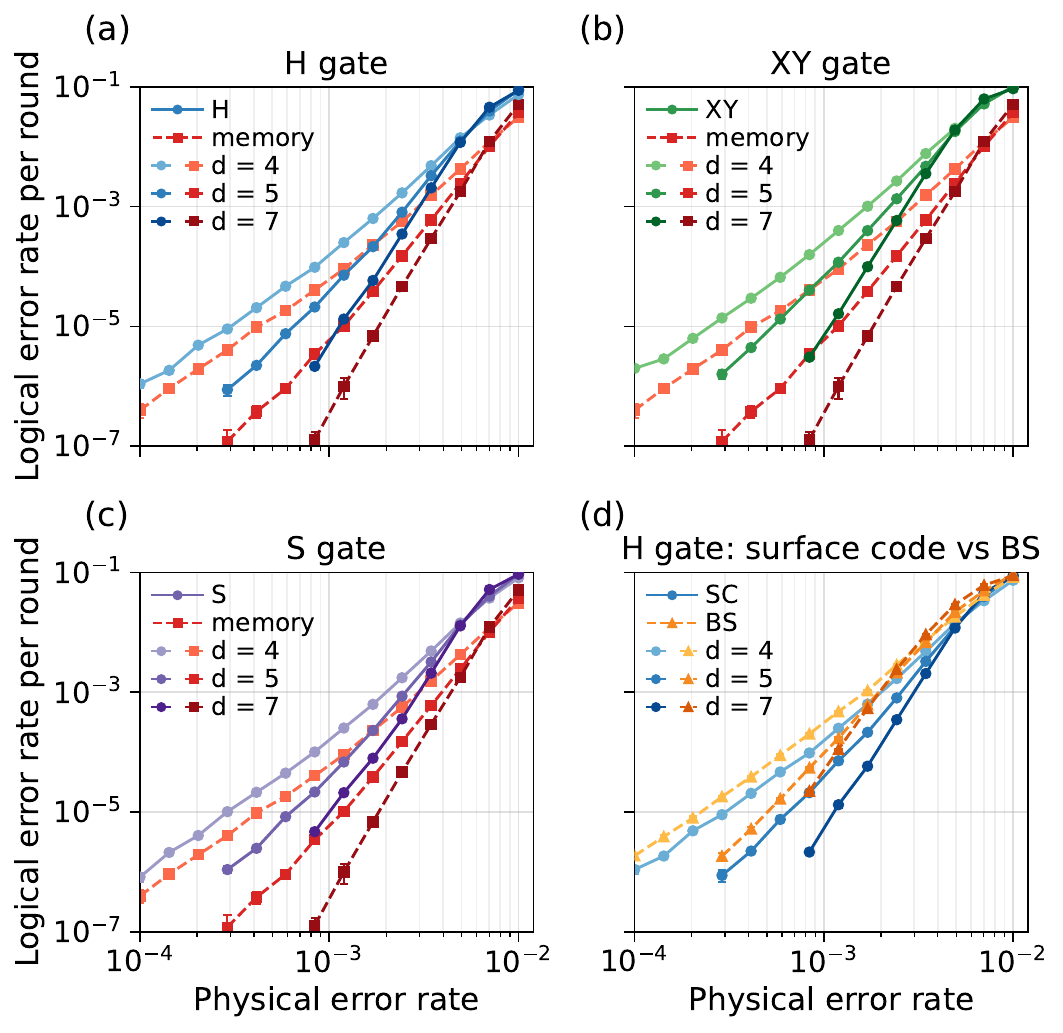}
    \caption{Quantum error correction simulations of logical gates implemented with the surface-code gadget on a $k=2$ La-cross code and for $d$ round of error correction: (a) Hadamard gate (blue) and (c) $S$ gate (violet) on a single logical qubit, and (b) $XY$ gate (green) on two logical qubits, each compared with a memory experiment (red). (d) comparison between the surface-code (blue) and Bacon--Shor (red) gadgets for the Hadamard gate. Darker shades correspond to larger code distances, $d = 4, 5, 7$. Benchmarked with the memory protocol, all gates (a)--(c) show a moderate increase of the logical error probability and decrease of the physical error threshold. A similar trend is observed in (d) for the Bacon-Shor gadget relative to the surface-code one. The logical error probability is normalized by the number of rounds and the error bars are standard deviations from the Monte Carlo simulations. The La-cross codes that we have simulated are $[[52,4,4]]$, $[[100,4,5]]$, and $[[202,4,7]]$. The BP+OSD decoder was used (see Methods for details).}
    \label{fig:fig3}
\end{figure}

To support the analytical claims of the previous sections, we show in Fig.~\ref{fig:fig3} the QEC simulations for the logical error probability of our protocols for the full Clifford logical gate set:  (a) Hadamard gate (blue), (b) $XY$ two-qubit gate (green) and (c) $S$ gate (violet) using the unrotated surface code gadget and each of them compared to a memory experiment (red) for the $k=2$ La-cross code family. In Fig.~\ref{fig:fig3}(d) we directly compare the logical error probability obtained with the surface code gadget (blue) and the Bacon-Shor code gadget (yellow) for the addressable Hadamard rotation. The three code instances shown are $[[52,4,4]]$, $[[100,4,5]]$, and $[[202,4,7]]$. 

We use \texttt{Stim}~\cite{Gidney_2021} to perform the QEC numerical simulations. For both logical rotation and memory, only one logical operator is considered. Motivated by experiments (see Methods), we choose a noise model that consists of two-qubit depolarizing errors for two-qubit gates and phase-flip errors for measurements and resets in the $X$-basis with equal error strength $p$. Single-qubit and idle-errors are neglected. For all gates, we employ a correlated decoding strategy~\cite{Zhou_2025,Cain2025,SerraPeralta2025} that allows us to insert a single round of error correction after each logical gate on both codes. For instance,for the Hadamard gate, we initialize the LDPC code in the $\ket{0}_L$ state by performing one round of error correction, apply another round between the two logical controlled gates and append the remaining $d-2$ rounds at the end for comparison with the memory experiments, where we perform $d$ rounds of QEC. Following the logical schemes of Fig.~\ref{fig:fig1}(f)--(g), we adopt the same approach for the other two logical gates. For the $S$ gate, the LDPC code is initialized in the $\ket{+}_L$ state, one round of error correction is performed between the $S$ gate on the auxiliary code and the transversal CZ, and the remaining $d-2$ rounds are appended at the end. For the $XY$ gate, both logical qubits of the LDPC code are initialized in $\ket{0}_L$, one round of error correction follows each of the two transversal CNOTs, and the remaining $d-3$ rounds are appended after the CZ, so that in all cases the total number of rounds is $d$. Data are normalized by the number of rounds as $P_L=1-(1-p_L)^{1/d}$, where $p_L$ is the ratio between the total number of errors and the number of shots. The error bars are standard deviations associated with the Monte Carlo simulations. The syndrome information is decoded using Belief Propagation with Ordered Statistics Decoder (BP+OSD) \cite{Roffe_2020,Roffe_LDPC_Python_tools_2022}. More details on circuit implementation and decoding are given in the Methods section.

For the surface-code gadget, our numerical results show a small decrease in the physical error threshold, from $p_{\rm th}^{\rm m} \approx 0.72\%$ for the memory to $p_{\rm th}^{\rm H} \approx 0.56\%$, $p_{\rm th}^{\rm S} \approx 0.55\%$ and $p_{\rm th}^{\rm XY} \approx 0.58\%$ for the Hadamard, $S$ and $XY$ gates, respectively, together with an increase in the logical error probability compared with the memory, due to the larger number of gate operations and hence of error locations. For the Hadamard gate, the Bacon--Shor gadget shows a further decrease in the physical error pseudothreshold, to $p_{\rm th}^{\rm H,BS} \approx 0.29\%$, i.e.about half of the surface-code value $p_{\rm th}^{\rm H} \approx 0.56\%$, together with a higher logical error probability. Crucially, the scaling of the logical error probabilities for the logical gates compared to the memory shows that our scheme is fault-tolerant -- i.e., the corresponding curves for the gates and memory are parallel and with the expected scaling for fault tolerance. For example, for the $d=4$ La-cross codes shown in Fig.~\ref{fig:fig3}, we find that the logical error probability correctly scales as $P_L\propto p^\nu$, with $\nu=\lceil d/2\rceil\approx2$ in the deep sub-threshold regime.

We verified the logical performance of the addressable single-qubit operations across all $K$ logical operators and found no differences within numerical precision. Consequently, the numerical simulations presented in Figs.~\ref{fig:fig3}(a), (c), and (d) correspond to the first logical operator of the La-cross code. For the entangling $XY$ operations, we selected two logical operators acting on two different rows of the same code block. No compilation optimization strategies were applied to ensure our protocol's evaluation reflects the worst-case scenario. 

This confirms both analytically and numerically that the protocols preserve the circuit-level distance of the code and are hence fault-tolerant. The numerical simulations also verify that our protocols perform the intended logical gates. For instance, for the Hadamard gate, the LDPC code is first initialized in its $\ket{0}_L$ state, rotated and then measured in the $X$ basis, yielding a deterministic result. Analogous checks are performed for the $S$ and $XY$ gates.

\subsection{Generalization to other quantum LDPC code families}
In this last section, we discuss what code structure is \emph{necessary and sufficient} for our logical gadget and discuss generalization to other quantum LDPC codes beyond La-cross codes. We first discuss the critical mathematical properties that a code must satisfy to benefit from our construction and after discuss what properties are instead desirable (although not strictly required) to facilitate near-term implementations with atomic qubits -- in particular neutral atoms -- via parallel atom shuttling.

Our protocol requires to find as many representatives of a logical operator as the code distance, $d$, trading the typical $\mathcal{O}(d)$ temporal redundancy of lattice surgery for an extra $\mathcal{O}(d)$ spatial redundancy. The motivation for this choice is that, although the total spacetime overhead remains $\mathcal{O}(d^2)$, gates -- especially transversal ones -- are typically a faster and cheaper resource than measurements in atomic platforms with long-range connectivity enhanced by atom transport. This extra fault-tolerant spatial redundification is precisely enabled by the possibility of finding $d$ equivalent representations of a distance-$d$ logical operator. 
Thus, the first necessary and sufficient condition for using our Bacon-Shor gadget is
\begin{center}
\emph{(i) Having as many representatives as the code distance for each pair of anticommuting logical operators $(X_L,Z_L)$.}    
\end{center}
This ensures that the logical CNOT and CZ required in our scheme are transversal and that the full protocol is fault-tolerant. If two equivalent representations overlapped on one data qubit, during the logical CNOT gate between the auxiliary gadget and the quantum LDPC code a $Z$ error on that data qubit could back-propagate into a pair of vertical $Z$ errors in the auxiliary gadget aligning with its $Z_L$ logical operator. We observe that changing the ordering of the representation pairing would not fix the problem, as two of such $Z$ errors can always be made aligning vertically with the $Z_L$ logical operator via multiplication by the $Z$ gauge operators.

In order to identify the generalized translation rules developed above for La-cross codes, the LDPC nature of the code is also important. The reason is that the ``low-density" property ensures that shifting a logical operator through stabilizer multiplication does not cause the weight of the resulting logical string to grow uncontrollably. More concretely, the necessary (though not sufficient) condition for a code to support $d$ non-overlapping equivalent representatives of the same logical operator is that $d^2\leq N$ (i.e., $d=\mathcal{O}(\sqrt{N})$ with pre-factor smaller than $1$). We observe that this is the optimal distance scaling of hypergraph product codes, including the constant rate quantum expander codes, where $K=\mathcal{O}(N)$ and $d=\mathcal{O}(\sqrt{N})$ \cite{xu2023constantoverhead}.

Our logical gadget sometimes requires pairing single-row representations of the auxiliary code logical operators with representations that span two or more rows of the quantum LDPC code lattice. This operation is fault-tolerant because it can only result in correlated errors that anti-align with the same-basis logical operator in the targeted code. This is the case because La-cross codes have open boundary conditions and therefore shortest representation of logical operators only have support on the main lattice of data qubits, and not on the secondary lattice. The necessary condition for our protocol to be fault-tolerant therefore is
\begin{center}
\emph{(ii-A) The targeted quantum LDPC code must have open boundary conditions}
\end{center}
or, rephrasing equivalently
\begin{center}
\emph{(ii-B) The targeted quantum LDPC code must support logical operators on only one lattice of data qubits, and not on both lattices.}
\end{center}
If this is not the case, the construction must be generalized, for example by gauging away half logical operators or by increasing the horizontal dimension--or both dimensions--of the logical gadget (e.g., $d\times d\rightarrow d\times2d$ or $2d\times2d$ if the code allows for at most two-row representations). We note that larger square configurations are preferable to rectangular ones due to their enhanced symmetry properties, namely the presence of larger automorphism groups. For instance, the protocol for the addressable $S$ gate discussed in this work no longer applies in the rectangular setting.

We let the ultimate problem of investigating whether our protocol admits generalizations to hypergraph product codes with non-cyclic seed codes, in particular constant rate codes, and of working out other example codes to future work. Instead, we discuss one last property that is desirable, although not required, in our scheme and that is automatically satisfied by all hypergraph product codes \cite{xu2023constantoverhead},
\begin{center}
\emph{($\circ$) The targeted quantum LDPC code has shortest logical operators which are single-row.} 
\end{center}
This property is not required because even if the equivalent logical operators are spread over the array of data qubits, provided they are not overlapping, each of them can be individually controlled by one row or column of the auxiliary code. Nevertheless, if they are single-row (or multiple-row but with ordered pattern), they can be easily implemented in neutral atom platforms via parallel atom transport with minimal number of movements (translations, dilatations and contractions), as we have commented for La-cross codes.

Finally, we observe that for constant or ultra-high rate quantum LDPC codes \cite{kasai2026breakingorthogonalitybarrierquantum,zhao2026ultrahighratequantumerrorcorrection}, for which the task of computing equivalent representatives of a logical operator is analytically non-trivial and numerically demanding, one could envision an architecture like the one in Ref.~\cite{cain2026shorsalgorithmpossible10000} that uses different quantum LDPC codes for different tasks. For example, one could then use constant-rate large LDPC codes for memory and smaller, computational efficient high- (but not constant-) rate LDPC codes for information processing via our auxiliary gadget as time-efficient alternative to code surgery.

\section{Discussion}

In conclusion, we have demonstrated a novel gate-based scheme for performing addressable quantum computation with high-rate quantum LDPC codes without relying on lattice-surgery-based techniques. We have shown how Clifford and non-Clifford resources can be fault-tolerantly prepared in an auxiliary Bacon-Shor or surface code, also using the possibility of morphing the Bacon-Shor code into a rotated surface code, where $|S\rangle$ states can be efficiently prepared and $|T\rangle$ states cultivated in a fold-transversal manner \cite{chen2024transversallogicalcliffordgates,sahay2025foldtransversalsurfacecodecultivation}. These universal resources can then be injected into a single logical qubit encoded in the quantum LDPC code via teleportation. Logical addressability is enabled by a careful qubit partition of the quantum LDPC code lattice consisting of the shortest $d$ non-overlapping equivalent representatives of the same $d$-long logical operator. Crucially, our protocol relies only on transversal controlled operations between the auxiliary and the quantum LDPC code, which can be realized in constant time. Therefore, our protocol outperforms all known lattice surgery schemes in terms of time overhead, while offering a comparable or better $\mathcal{O}(d^2)$ spacetime overhead overall. Additionally, since we do not require measuring all the stabilizers of the auxiliary code at each QEC round, the total gate count of our scheme is also substantially reduced. 

The protocol we have presented in this work is hardware-agnostic, and hence suitable for implementation in several qubit platforms. Rydberg atom and trapped-ion platforms are promising for near-term experimental realizations as they display native long-range couplings and flexible connectivity due to atom shuttling \cite{Bluvstein_2023}, atom bus \cite{bergonzoni2026faulttolerantquantumcomputationstatic} and long-range Rydberg and Coulomb interactions \cite{pecorari2024highratequantumldpccodes,saffman2025quantumcomputingatomicqubit}, respectively. In addition, atomic qubits typically suffer from slow measurements. Therefore, fast schemes that do not rely on lattice surgery are  particularly useful to enable error-corrected quantum computation. 

In this work, we focus on La-cross quantum LDPC codes because their simple code structure allows us to identify the translation rules among individual logical operators. This, in turn, enables the development of a gate-based, fault-tolerant protocol that can operate on individual logical qubits in an addressable manner. We have discussed the necessary and sufficient conditions for our protocol to extend to other quantum LDPC codes in a fault-tolerant manner, namely finding $d$ equivalent non-overlapping representatives of the same logical operator and having logical operators localized on only one lattice of data qubits, which can be realized via gauging or naturally with open boundary conditions. It remains an interesting open question whether our protocol can be generalized to other high-rate or constant-rate quantum LDPC code families, such as, specifically, bivariate bicycle codes \cite{Bravyi2024}, or lifted product codes \cite{lpcodes,xu2023constantoverhead,Old_2024}, which share similar stabilizer and logical operator structure.

Another important open direction is the development of fast decoding algorithms for quantum LDPC codes. In this work, we employed the BP+OSD decoder, which currently represents the state of the art in quantum LDPC decoding. Although highly accurate and therefore appropriate for benchmarking, the computational complexity of the OSD part can scale as $\mathcal{O}(N^3)$, resulting in a huge time overhead already for moderate system sizes, $N$. As a result, BP+OSD becomes computationally prohibitive for applications involving deep quantum circuits. Our logical gadget relies on gate teleportation, and while this poses no issue for preparing Clifford resources, implementing non-Clifford gates requires real-time decoding to avoid the accumulation of unprocessed error syndromes. This can result in a backlog that would quickly degrade the protocol’s performance. A variety of decoders have emerged recently, many of which augment belief propagation with novel post-processing techniques to bypass the computational limitations of ordered statistics decoding \cite{Hillmann_2025,10619083,iolius2025almostlineartimedecodingalgorithm}. Parallel efforts have focused on generalizing matching-based decoders to quantum LDPC codes via over-matching, which resolves space-like hyperedges without sacrificing the inherent speed of matching algorithms \cite{sahay2026matchingdecoderbivariatebicycle}. This paradigm constitutes a highly promising strategy for enabling fast, correlated decoding across our logical gadget, analogous to the matching-based approaches applied to decode across transversal gates between surface codes \cite{SerraPeralta2025,Cain2025}.

\section{Acknowledgements}
We gratefully acknowledge discussions with Gavin K. Brennen. This research has received funding from the European Union’s Horizon 2020 research and innovation programme under the Horizon Europe programme HORIZON-CL4-2021-DIGITAL-EMERGING-01-30 via the project 101070144 (EuRyQa) and under the Marie Sk\l{}odowska-Curie grant agreement number 101120240 (MLQ) and from the French National Research Agency under the Investments of the Future Program projects ANR-21-ESRE-0032 (aQCess), ANR-22-CE47-0013-02 (CLIMAQS), ANR-17-EURE-0024 (QMat), and ANR-22-CMAS-0001 France 2030 (QuanTEdu-France). Computing time was provided by the High-Performance Computing Center of the University of Strasbourg. Part of the computing resources were funded by the Equipex Equip@Meso project (Programme Investissements d'Avenir) and the CPER Alsacalcul/Big Data.

\emph{During the final stages of this work, a study appeared \cite{xu2025batchedhighratelogicaloperations}, which  develops a different approach to perform high-rate addressable Clifford gates in CSS quantum LDPC codes.}

\bibliography{ref}

\section{Methods}
\subsection{Deriving equivalent representations of a logical operator: an algebraic approach}
In the following, we describe how to derive the different equivalent representations associated with a logical operator $X_L$ of arbitrary $k$ La-cross codes. The discussion is analogous for $Z_L$ operators.

(\emph{i}) \emph{Derive the shortest logical operator.}
Starting from the matrix $H_X$ in Eq.(\ref{eq:hgp_matrix}), the logical operators are obtained by finding the vectors $\mathbf{x}$ that solve the problem $H_X\,\mathbf{x}=0$. 
The vector $\mathbf{x}$ represents the support of the logical operator $X_L$ acting on $N=n^2+r^2$, with $r=n-k$, physical qubits:
    \begin{align}
        X_L = \bigotimes_{l=1}^{N} X_l^{\mathbf{x}_l},
    \end{align}
where $X_l$ is a Pauli $X$ operator acting on the $l$th qubit, $\mathbf{x}_l$ is the $l$th component of the vector $\mathbf{x}$, and hence $X_l^{\mathbf{x}_l}$ is an operator acting at the qubit location specified by $\mathbf{x}_l$.
The physical qubits of the array are indexed in increasing order from left to right and from top to bottom, starting from the $n \times n$ main lattice and then the $r \times r$ sublattice. We solve the problem numerically via Gauss elimination and select the shortest solutions defined on a single row of the main lattice. 
There exist $k^2$ independent solutions that represent the different inequivalent logical operators.

(\emph{ii}) \emph{Finding equivalent representations.}
Let us consider a particular solution $X_L$ that has support on the row $i$ of the main lattice, characterized by the support pattern $P =\{p|\mathbf{x}_{(i-1)n+p} = 1\}_{1\leq p \leq n} $, which in the following we index as $X_L^{(i,P)}$ and derive its different equivalent representations.
Any logical operator $X_L^{(i,P)}$  can be translated vertically along the main lattice via multiplication by $X$-stabilizers.
The logical operator $X_L^{(i,P)}$ on row $i\leq r$, can be translated into a two-row equivalent representation acting on the two underlying $(i+1)$th and $(i+k)$th rows as follows.
Consider the set of $X$-stabilizers corresponding to the rows of $H_X$ which we re-index as $\{S^X_{i,p}:=S^X_{(i-1)n+p}\}_{\tiny{\substack{i=1,...,r \\ p=1,...,n}}}$.
One can show that applying the subset $\{S^X_{i,p}\}_{p\in P}$ of $X$-stabilizers, whose ancilla qubits lie in the row immediately below the data qubits in the support $P$ of $X_L^{(i,P)}$, and that couple upward to them via nearest-neighbor coupling yields the desired translation
\begin{align}
\label{eq:operator_translation}
\prod_{p \in P} S^X_{i,p} \, X_L^{(i,P)} = X_L^{(i+1,P)} \, X_L^{(i+k,P)}, 
\qquad i \leq r .
\end{align}
We note that this relation is symmetric under the permutation of the indices of the logical operators, that is, one can perform vertical translations both upward and downward. Therefore, any two-row representation acting on any given pair of rows chosen among the three rows $i$, $i+1$, or $i+k$ can be mapped into a single-row representation acting solely on the row that is not part of the pair.

We introduce a convenient approach to perform these calculations directly on the vectors $\mathbf{x}^{(i,P)}$: 
We construct the set of vectors $\{h^X_{i,p}:=h^X_{(i-1)n+p}\}_{\tiny{\substack{i=1,...,r \\ p=1,...,n}}}\in\mathbb{F}_2^{N}$ that correspond to the $nr$ $X$-stabilizers. The product of stabilizers in Eq.~\eqref{eq:operator_translation} can be reexpressed as a macro-stabilizer vector, that is, a sum of stabilizer vectors, $M^X_{(i,P)}$, defined as
\begin{align}
    \label{eq:matrix_stab}
 M^X_{(i,P)} = \sum_{p\in P}\, h^X_{{i,p}}\;.
\end{align}
With this new notation, Eq.~(\ref{eq:operator_translation}) reads as a sum of the vectors $M^X_{(i,P)}$ and $\mathbf{x}^{(i,P)}$, 
\begin{align}
    \label{eq:matrix_action}
    M^X_{(i,P)} +\,\mathbf{x}^{(i,P)} = \mathbf{x}^{(i+1,P)} + \mathbf{x}^{(i+k,P)}.
\end{align}
By varying $i$, it is possible to construct $r$ sums of stabilizer vectors $\{M^X_{(i,P)}\}_{i=1}^r$, so that all possible translations between the rows are obtained via all possible combinations of these sums. To obtain as many representations as the length of the considered logical operator, our protocol requires taking all the possible combinations of $M_{(i,P)}^X$ acting on $X_L^{(i,P)}$ to identify all the equivalent logical operators. Then, these operators are ordered according to their length, and, finally, we retain only the shortest non-overlapping representatives.

For the logical $Z$ operators, the discussion is analogous, but applied to the columns of the lattice.

We find by inspection that for $k=2,3$ La-cross codes it is always possible to find as many equivalent representations of the same logical operator as its length which span either a single row or two rows of the main lattice. Instead, for $k=4$ we find lattice sizes $n$ for which some equivalent representatives span three rows. It follows that for our logical gadget to be fault-tolerant with high-rate La-cross codes, $\ell\leq k-1$-long representations of a logical operator must be considered. The protocol remains valid, but requires more steps per logical operation. For an optimal compilation strategy, it may be advantageous to restrict to code sizes that admit small-$\ell$ representations for large-$k$ La-cross codes (e.g., $\ell=2$ for the case $k=4$).

Finally, we observe that for some values of the parameters $k$ and $n$ (the linear system size), we found by inspection some rare cases where the pair $(X_L,Z_L)$ of anticommuting logical operators have lengths that differ by one unit, that is, for example, $d_X=d+1$ and $d_Z=d$. These ``rectangular" configurations offer a slightly asymmetric protection against $X$ and $Z$ errors because of the two different code distances ($d_X\neq d_Z$). Nevertheless, since computing logically requires mixing the $X$ and $Z$ basis, the overall protection offered by the code is set by the shortest distance, i.e., $d_Z=d$ in this example. Therefore, we argue that these cases can still be handled with our logical gadget using an auxiliary $d\times d$ Bacon-Shor or surface code without loss of fault tolerance or generality.

\subsection{Details on quantum error correction simulations}
In the following, we provide additional details about the QEC numerical simulations of the logical overcomplete Clifford set of gates discussed in the main text. Since the general structures of the protocols has already been described, we focus here primarily on the decoding strategy and noise model employed.

The logical transversal property of our scheme enables the use of a correlated decoding strategy~\cite{Zhou_2025,Cain2025,SerraPeralta2025} that significantly reduces the number of error correction cycle required. Indeed, only a single round of error correction is needed after a logical gate as errors propagate deterministically through Clifford gates and can therefore be tracked in software rather than corrected physically at each step.
To implement this decoding strategy, we use the \texttt{detectors} in \texttt{Stim}, defined as the binary addition of measurement outcomes. In a standard quantum memory experiment, one typically associates a detector to each stabilizer by pairing its measurement outcomes across two consecutive rounds. The resulting detector signal is therefore, only sensitive to changes in the stabilizer value between rounds that corresponds to actual physical or measurement errors. In a logical circuit, the logical gates propagate errors between code blocks. To avoid flagging the same error multiple times, detectors must be designed to be sensitive only to newly introduced errors. This is achieved by backpropagating each stabilizer through the preceding logical gate and rewriting it as a product of stabilizers in the updated frame. This is always possible thanks to the fault-tolerance of our gadgets (see main text). Detectors are then defined as the parity of the measurement outcomes associated with these stabilizers. The explicit backpropagation rules for all stabilizer types and codes are given in the main text. 

The final measurement of the data qubits in both codes provides additional information that can be exploited to improve decoding performance. When the data qubits are measured in a certain basis, one can reconstruct the value of stabilizers of the corresponding type by properly combining the measurement outcomes. These “artificial” stabilizers can then be backpropagated to the previous error correction round, in the same way as ordinary stabilizers, to define additional detectors. For the LDPC code of the Hadamard and XY gates, the end of a circuit corresponds to a memory experiment and this backpropagation is trivial.The S gate requires an additional step. Since the LDPC code is initialized in the $\ket{+}_L$ state, the S gate leaves it in $S\ket{+}_L = \ket{+i}_L$, which is not an eigenstate of the logical $X$ operator, so that a final $X$-measurement would yield a random outcome even in the absence of noise. To obtain a deterministic outcome, we append a noiseless fold-transversal S$^\dagger$ gate on the LDPC code after the last round of error correction, which maps $\ket{+i}_L$ back onto $\ket{+}_L$ before the destructive $X$-measurement. This operation is not part of the gate protocol: it serves only to verify the logical action of the S gate, and it is taken to be noiseless so as not to add further error locations. Since the S$^\dagger$ maps each $X$-stabilizer of the LDPC code onto its product with the mirror $Z$-stabilizer, each final detector compares the reconstructed $X$-stabilizer with the last measurements of both the $X$-stabilizer and its mirror $Z$-stabilizer. In contrast, for the auxiliary code, the $X$-measurement is performed immediately after the transversal CZ operation. The backpropagated “artificial” $X$-stabilizer is equal to the product of a $Z$-stabilizer of the LDPC code and an $X$-stabilizer of the auxiliary code. Since this product is never measured in the circuit, the corresponding detector is defined as the combination of the last measurements of its factors before the CZ.

In addition to the detectors, we must define a \texttt{logical\_observable}, that is typically the logical measurement outcome of a logical qubit. In the noiseless situation, this observable should be deterministic. When errors occur, the value of this observable might be affected and the decoder's role is to infer the noiseless logical observable value based on detector outcomes. In our scheme, logical measurements in the auxiliary and in the LDPC code are always random when considered separately, but their joint product is deterministic. We, therefore, define the latter as the logical observable of our system in order to evaluate the logical error of our protocol. For the S gate, this holds thanks to the noiseless S$^\dagger$ introduced above, so that the observable is again the product of the logical $X$ outcome of the LDPC code and that of the auxiliary code.

We used a circuit-level noise model that considers errors only after two-qubit gates, measurements, and resets. Single-qubit gates and idle operations are assumed to be perfect. That is motivated by the fact that these operations typically have one order of magnitude lower infidelities than two-qubit gates and measurements in most quantum computing platforms, such as neutral atoms. For two-qubit gates we enforced two-qubit depolarizing noise, that is, we assumed uniformly distributed Pauli errors drawn at random from $\{I, X, Z, Y \}^{\otimes2}\backslash\{I \otimes I\}$ with probability $p/15$. For measurements and resets in the $X$-basis, we inject phase-flip errors with probability $p$.

From the set of detectors, the logical observable and the noise model defined above, \texttt{Stim} is able to derive a detector error model (DEM) i.e. the set of all single error mechanisms along with their associated probabilities, the flipped detectors and potentially the flipped logical observable. We decoded the syndrome information using Belief Propagation with Ordered Statistics Decoder
(BP+OSD) \cite{Roffe_2020,Roffe_LDPC_Python_tools_2022}. We used the minimum-sum variant of Belief Propagation, setting the number of iterations to $4$ and optimizing the scaling factor, $s$, to obtain the lowest logical error probabilities and the correct asymptotic scaling without error floors. The optimal scaling factors were found to be $s=0.2$ for the logical gate protocols and $s=0.3$ for the memory experiment. OSD was used in combination sweep mode up to order $1$. Monte Carlo samplings were performed using the Sinter library.

We note that for the logical gate protocols we have used BP+OSD over the full decoding problem, consisting of the two code memories -- BS or SC and La-cross -- along with the transversal entangling operations between them. That is because the DEM of quantum LDPC codes naturally suffers from hyperedges \cite{Roffe_2020} and the logical entangling operations between different codes introduce additional hyperedges to the decoding graph due to errors propagating from one code to the other. Although this is a legitimate decoding strategy, hypergraph complexity and time overhead increase, which can ultimately lead to under-performance of the decoder. To address this issue, more efficient decoding techniques consisting in decomposing the decoding problem into different parts have been introduced \cite{puriPRXQuantum.6.020326}. We let to future work the issue of addressing the decoding problem for our BS gadget with more refined and possibly more efficient methods.

\begin{figure}
    \centering
    \includegraphics[width=\linewidth]{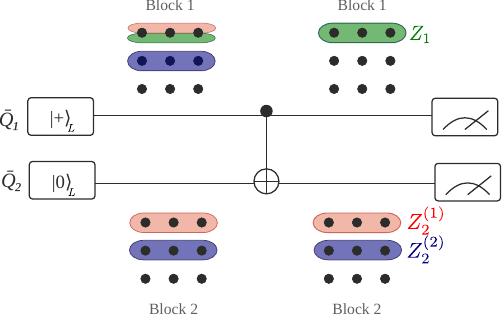}
    \caption{A Bell circuit involving two surface code patches $\bar{Q}_1$ and $\bar{Q}_2$. The logical observable $Z_1Z_2$ is reliable for this circuit. Depending on the representation chosen for $Z_2$ (red or blue), its backpropagation may not overlap with $Z_1$ (green), hence resulting in a non trivial overlap with non-deterministic stabilizers spoiling the reliability of $Z_1Z_2$.}
    \label{fig:correlated_decoding}
\end{figure}

\subsection{Pauli frame tracking in correlated decoding}
In this section, we provide additional details regarding the accurate tracking of the Pauli frame during correlated decoding within our protocol, and explicitly demonstrate how to pair the logical operator representatives of the Bacon-Shor or surface code and the target quantum LDPC code to prevent sign inconsistencies.

We recall that correlated decoding consists in decoding only reliable logical observables, namely those that can be backpropagated and terminated at a reset in the same logical basis. Here, we highlight that correlated decoding relies implicitly on a consistent choice of logical representatives across blocks \cite{SerraPeralta2025}. When this consistency is broken, the construction can still be salvaged by absorbing additional detectors into the reliable observable. This generalization is essential for the our protocol, where a consistent choice between the auxiliary code block and the quatum LDPC code block is not obvious a priori. To facilitate the discussion and build intuition, we detail this mechanism using the surface code, which is the simplest quantum LDPC code in our considered family ($k=1$). However, the exact framework generalizes straightforwardly to La-cross codes as well.

Consider a generic logical circuit with transversal entangling CNOT gates between surface codes. Each surface code block encodes a logical qubit, and one can pick the same row/column (e.g. the first) as the logical operator representative for each logical measurement \cite{SerraPeralta2025}. With this convention, a logical operator backpropagating into a neighboring block cancels exactly against that block's representative. Choosing a different representative in one block breaks the cancellation: the two representatives now differ by a product of stabilizers, and when this stabilizer product is backpropagated further it may not commute with logical resets. The resulting observable then picks up non-deterministic detectors that correspond ultimately to the Pauli frame at those resets and ceases to be reliable — purely as an artifact of the initial choice of representative.

In Fig.~\ref{fig:correlated_decoding} we illustrate the problem for a simple Bell circuit comprising a single transversal CNOT gate. Consider two surface code blocks $1$ and $2$: $Z_1 Z_2$ is a reliable logical observable for this circuit with the representation of $Z_1$ chosen to be the data qubits on the first row. If $Z_2$ is chosen consistently, (i.e. same row as $Z_1$ but on block 2), its backpropagation through the transversal CNOT will exactly overlap with $Z_1$ and their combined contribution in the logical observable $Z_1Z_2$ does not overlap with the stabilizers of the reset $\ket{+}_L$. On the other hand, if the representation of $Z_2$ lies on another row, the combination of both logical measurements will result in a non trivial overlap with the non-deterministic $Z$ stabilizers of the reset $\ket{+}_L$ operation.

Fortunately, reliability can be restored systematically by including precisely these non-deterministic stabilizers into the definition of the observable. This generalizes the construction of the original correlated decoding proposal in Ref.~\cite{SerraPeralta2025} by removing the constraint that all logical qubits share a consistent representative. While the generalization is not required in the pure surface code case — where a consistent choice always exists — it becomes essential for arbitrary La-cross codes. This is because the appropriate alignment between auxiliary and La-cross code representatives is not obvious and depends on the implementation details of the transversal CNOT and CZ gates. Our construction therefore allows any representative to be selected as the reliable logical observable for each logical measurement, with the corresponding non-deterministic reset stabilizers absorbed systematically to preserve reliability.

\begin{figure}[t!]
    \centering
    \includegraphics[width=\linewidth]{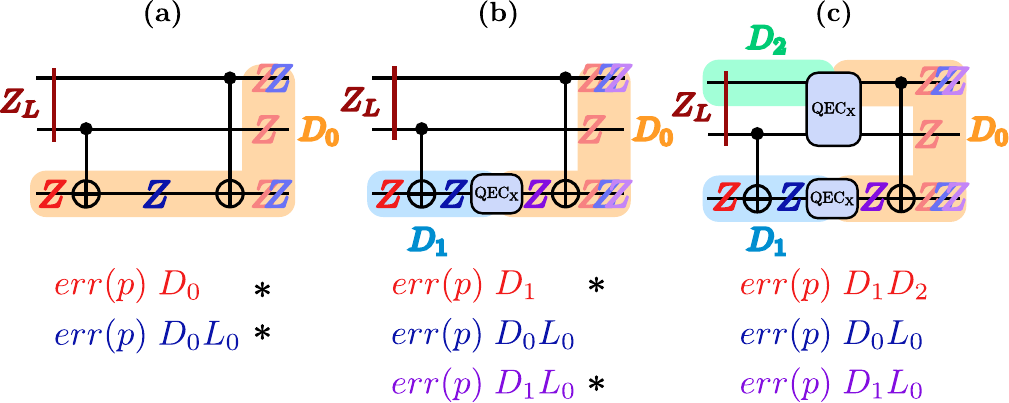}
    \caption{Two physical CNOT controlled by the Bacon-Shor and targeting twice the same data qubit in the quantum LDPC code (assuming overlapping representatives). Errors at different circuit locations are drawn with different colors and so are their corresponding the detector error model entries. Errors propagate thought the circuit from left to right (color grading). Detectors and detecting regions are highlighted with colors. (a) The protocol is not fault-tolerant because different low-weight error mechanisms can fire the same set of detectors with opposite logical outcomes. (b) Upon inserting one $X$ stabilizer measurement round (QEC$_X$) in the LDPC code, the circuit is still not fault-tolerant. (c) Inserting a QEC$_X$ round in both Bacon-Shor and LDPC code in between the two physical CNOT succeeds in resolving the pair of correlated $Z$ errors in a fault-tolerant manner, but results in a $\mathcal{O}(d)$ asymptotic circuit depth (because we now measure nonlocal stabilizers in the Bacon-Shor code) and in the loss of an asymptotic threshold for our gadget. Similarly, the surface code gadget fails because of the impossibility of performing QEC rounds and simultaneously maintaining the sub-lattice completion that makes the gadget possible.}
    \label{fig:overlapping}
\end{figure}

\subsection{Overlapping representatives}
In the Results section, we identified that a code allows for logical computation using our logical gadget if and only if it admits $d$ equivalent, non-overlapping representatives. In the following, we explain in more detail why the non-overlap assumption is critical for the protocol to work in a fault-tolerant manner.

Let us suppose that two equivalent representation of the same $X_L$ logical operator overlap on one data qubit in the quantum LDPC code lattice. When pairing the auxiliary code and the LDPC code via a logical CNOT, that data qubit participates in two physical CNOT gates. Consequently, if a $Z$ error is present, it can back-propagate through the CNOT into two correlated $Z$ errors in the auxiliary code that align vertically (modulo multiplication by $Z$ gauge operators). Pairs of vertically aligning $Z$ errors are hook error string that align with the vertical $Z_L$ logical operator of the Bacon-Shor code, which results in a halved circuit-level distance of $d/2$ instead of $d$, violating fault tolerance. This criticality is also revealed by the detector error model of the circuit [Fig.~\ref{fig:overlapping}(a)]. Let $D_0$ be the detector checking the parity of the stabilizer $X$-measurements of the auxiliary and the LDPC code, a $Z$ errors occurring before the first CNOT fires the detector but flips twice the logical operator, resulting in no visible observable flip. Instead, a $Z$ error occurring in between the two CNOT gates fires the detector and flips once the logical operator resulting in a logical change of frame. In the detector error model, these two error mechanisms fire the same set of detectors with opposite logical outcomes (equivalently, their sum produces a logical change of frame with trivial detector syndrome) and are hence symptoms of a non-fault-tolerant circuit.

To mitigate this issue, one could insert one round of $X$ stabilizer measurements in the LDPC code in between the two physical CNOT gates. In this way, the two error mechanisms described above have different distinguishable detector syndromes, namely $D_1$ and $D_0L_0$ according to the notation defined in Fig.~\ref{fig:overlapping}(b). However, if a $Z$ error happens in between the two CNOT gates but before the QEC round, it would fire the first detector, the second detector stays flipped and does not record any parity change, and the logical observable gets flipped once, resulting in $D_1L_0$. Since two distinct error mechanisms with syndromes $D_1$ and $D_1L_0$ are possible, the circuit is still not fault-tolerant.

In order to solve this distinguishability problem, one would need to insert a $X$ stabilizer measurement round on the auxiliary code as well. In this way, $Z$ errors at any circuit location fire distinct sets of detectors, allowing to resolve the pair of correlated errors while preserving fault tolerance. However, this is not possible for any of the gadgets but for two distinct reasons. For the BS gadget, since we initialize it in the $Z$ gauge and $X$ stabilizers are nonlocal ($X$ and $Z$ gauge operators do not commute), measuring nonlocal operators is problematic to ensure an asymptotic threshold for our protocol in the limit $d\rightarrow\infty$. Therefore, while still allowing for a finite pseudothreshold at finite system size, the condition of $d$ equivalent non-overlapping representatives is critical for the fault tolerance and scalability of the Bacon-Shor gadget.

For the SC gadget, it is not possible to insert an intermediate round of stabilizer during the transversal operation since a compatible completion of the stabilizers on the secondary lattice would fail.

\section*{DATA AVAILABILITY}
The code reproducing the error correction simulations shown in this work can be made available upon request to the authors.

\section*{AUTHOR CONTRIBUTIONS}
L.P., F.P.G., and H.P. conceived the logical gate scheme and developed the numerical simulation code. F.P.G. performed the error correction simulations. G.P. supervised the project. All authors discussed the results and contributed to writing the manuscript.

\section*{COMPETING INTERESTS}
G.P. is co-founder and shareholder of QPerfect, H.P. is an employee at QPerfect. The remaining authors declare no competing interests.

\end{document}